\documentclass{aastex62}

\graphicspath{{./}{figures/}}
\usepackage{physics}

\received{January 1, 2018}
\revised{January 7, 2018}
\accepted{\today}
\submitjournal{ApJ Lett}

\shorttitle{ISM Fe oxidation state}
\shortauthors{Westphal et al.}

\begin{document}

\title{Measurement of the Oxidation state of Fe in the ISM using X-ray Absorption Spectroscopy}

\correspondingauthor{Andrew J. Westphal}
\email{westphal@berkeley.edu}

\author[0000-0003-1156-3776]{Andrew J. Westphal}
\affil{Space Sciences Laboratory, U. C. Berkeley}

\author{Anna L. Butterworth}
\affiliation{Space Sciences Laboratory, U. C. Berkeley}

\author{John A. Tomsick}
\affiliation{Space Sciences Laboratory, U. C. Berkeley}

\author{Zack Gainsforth}
\affiliation{Space Sciences Laboratory, U. C. Berkeley}



\begin{abstract}

	The oxidation state of iron in the interstellar medium (ISM) can provide  constraints on the processes that operated on material in the protosolar disk.  We used synchrotron-based X-ray absorption spectra of several mineral standards and two kinds of primitive extraterrestrial materials to constrain the oxidation state and mineralogy of the host-phase of ISM Fe as measured by X-ray observations of Fe-L ISM absorption from the {\em Chandra X-ray Observatory}.  Given the initial oxidation state of the building blocks of the Solar System, we explore nebular and parent-body processes that eventually led to the remarkable diversity of oxidation states of primitive Solar System materials.   Oxidation of cometary material appears to have taken place in the nebula, before incorporation into cometary nuclei, although the mechanism is unknown. 

\end{abstract}



\section{Introduction} \label{sec:intro}

The most primitive materials in the Solar System reside in comets. The size distribution of minerals in even the finest-grained components of cometary dust particles, however, is inconsistent with the size distribution of ISM dust as inferred from astronomical observations \citep{gainsforth2017,draine2009}. This simple observation leads to the conclusion that some processing of material must have occurred in the protosolar disk, even in the outer comet-forming region,  in the epoch before comets formed. 

The nature of that processing is unknown, but the oxidation state of Fe  can provide important constraints.
Processes and environments that modify materials in the nebula can either oxidize Fe, for example by heating in the presence of water vapor \citep{grossman2012}, or reduce it,  for example through ion bombardment \citep{carrez2002}.  The oxidation state of Fe varies widely among solar system materials.  Indeed, Fe oxidation state was used by  \citet{ureycraig1953} to define the first rigorous, quantitative classification scheme for meteorites.  So the oxidation state of Fe in the ISM is an important piece of the puzzle, because it defines the starting chemical state of the protosolar cloud that eventually formed the Solar System.     

Fe also plays a special role in probing the relative rates of dust destruction and formation in the ISM.
While other major rock-forming refractory elements (Mg, Si) are synthesized principally in stars with dust-forming stellar outflows,  
Fe is synthesized mostly in Type Ia supernovae, which do not produce dust-forming outflows.    At least 70\% of freshly-synthesized
Fe is injected into the interstellar medium in the gas phase \citep{dwek2016}.  The astronomical observation that $>$90\% of Fe is depleted from the gas-phase in the ISM \citep{jenkins} leads
to the  conclusion that Fe must  condense in the ISM.    The condensation mechanism and host phase(s) are unknown.
Circumstantial evidence however points away from co-condensation of Fe with Mg and Si, pointing toward a non-silicate host for Fe in the solid phase \citep{dwek2016}.

We have combined synchrotron-based X-ray absorption spectra of several mineral standards and two kinds of primitive extraterrestrial materials with X-ray observations of Fe-L ISM absorption observations from the {\em Chandra X-ray Observatory} \citep{weisskopf2002} to constrain the mineralogy of the host-phase and oxidation state of ISM Fe.

\section{ISM observations} \label{sec:ismobservations}

We combined six {\em Chandra} observations (Table 1) of the X-ray binary Cyg X-1, with a total exposure time of 119 ks.  We used observations for which the X-ray source was to the side of its binary companion so as to avoid any possible contamination within the Fe spectrum from the stellar wind, and the orbital phases of Cyg X-1 during the observations are provided in Table 1.  The observations were made with the High Energy Transmission Grating (HETG; \citealt{canizares2005}), and we used the data from the Medium Energy Grating (MEG), downloading the spectra from the TGCat website \citep{huenemoerder2011} and processing with the Sherpa software package \citep{Freeman}. Final spectra combine the MEG+1 and MEG-1 diffraction orders.  The data include the K absorption edge of singly-ionized Ne, nominally at 848.6 eV.    The position of the Ne absorption line is consistent with an energy blueshift of 0.2$\pm$0.1 eV, and we included this blueshift when fitting the Fe-L absorption edges.

\begin{table}
        \centering
        \newcommand{\TableWidth}{0.55\textwidth}
\begin{tabular*}{\TableWidth}{ccccc}
        ObsID & T$_0$$^1$ & Orbital Phase ($\phi$)$^2$ & $\Delta\phi$$^3$ & Exposure (ks) \\
        \hline
        107   & 51470.80  & 0.741        & 0.034           & 16.450 \\
        1511  & 51555.30  & 0.836        & 0.030           & 14.515 \\
        2741  & 52302.20  & 0.204        & 0.014           & 6.774 \\
        2743  & 52377.87  & 0.711        & 0.014           & 6.774 \\
        3815  & 52702.66  & 0.765        & 0.124           & 59.994 \\
        13219 & 55597.27  & 0.635        & 0.030           & 14.515 \\
        \hline
        \multicolumn{5}{p{\TableWidth}}{$^1$Modified Julian date at beginning of observation.} \\
		\multicolumn{5}{p{\TableWidth}}{$^2$ Orbital phase of Cyg X-1 at the midpoint of the observation (phase zero corresponds to the black hole being behind the supergiant companion.)} \\
\multicolumn{5}{p{\TableWidth}}{$^3$ Change in orbital phase from the start to the end of the observation (in units of 5.599829 terrestrial days).} \\
\end{tabular*}
\caption{{\em Chandra} observations of Cygnus X-1 used for this analysis.}
\end{table}

\section{Acquisition of absorption spectra} \label{sec:absspectra}

We selected Fe-bearing materials representing various oxidation states, relevant formation processes, and Fe site symmetries, many previously characterized during other projects. Reduced Fe samples included Fe metal and sulfides, such as troilite (FeS, formed by metal and S gas) and ferromagnetic pyrrhotite (Fe$_{7}$S$_{8}$). We selected common, naturally occurring Fe oxide/silicate compounds found in primitive solar system materials. Samples included oxides of mixed Fe$^{2+}$/Fe$^{3+}$ in chromite and Fe$^{2+}$ in magnesiospinel.  We measured Fe$^{2+}$ in crystalline (Fo$_{68}$ olivine), and Fe$^{2+}$ to Fe$^{3+}$ in amorphous silicates including a natural volcanic glass, Glass with Metals and Embedded Sulfides (GEMS) in an interplanetary dust particle (RB-12A31-2), and fine-grained material associated with a large sulfide recently identified in the Stardust collection from comet Wild 2 \citep{butterworth2010,gainsforth2015,stodolna2013}.  

\begin{figure}
 \vspace*{-0.0in}
\includegraphics[width=7in]{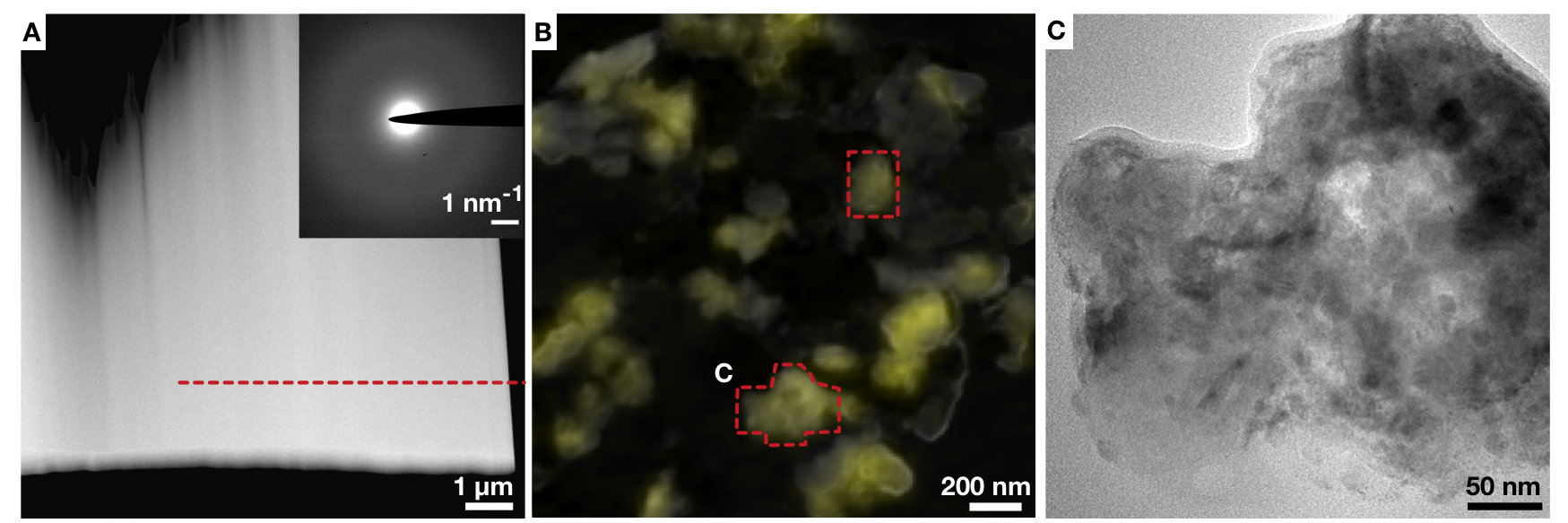}
 \vspace*{-0in}
 \caption{Sample selections for acquiring Fe L-edge spectra from Fe-bearing amorphous silicate and GEMS. A) TEM HAADF image of a 300-nm thick section of reticulite glass without inclusions, prepared by Focused Ion Beam (FIB). STXM line spectra were acquired at the position shown by the dashed red line.  Inset, TEM electron diffraction of FIB section.  The lack of diffraction spots confirmed that the glass was amorphous. B) Chondritic porous interplanetary dust particle with GEMS (red outlines). The TEM HAADF image overlaid with STXM Fe map (yellow) extracted from a stack of 120 images at energies across the Fe L-edge (680 to 750 eV). Spectra were integrated within each dashed red region. C) TEM bright field image of selected GEMS, showing low density amorphous mass with dark Fe metal and sulfide nm-sized inclusions.}
\label{fig:STXM}
\end{figure}

We used an ultramicrotome to cut $\sim$100-nm thick sections of samples onto Cu transmission electron microscope (TEM) grids, or prepared sections using an FEI DualBeam focused ion beam (FIB) at the Molecular Foundry (MF) at Lawrence Berkeley National Laboratory (LBL) .  We identified phases and determined their crystal/amorphous character using an FEI Titan TEM at the MF at LBL  (Fig. \ref{fig:STXM})

We collected high spatial resolution (50 nm) X-ray absorption stacks at beamline 11.0.2.2 at the Advanced Light Source at LBL.  Stacks were acquired as images at successive energies across the Fe L$_{23}$-edges.  The highest energy resolution was 100\,meV. We converted stack data to Optical Density (OD) XANES spectra by aligning the images and normalizing counts to the incident beam intensity using aXis2000 software (http://unicorn.mcmaster.ca/aXis2000.html).
We converted OD to cross-section spectra in MATLAB by first subtracting the background using a linear fit to the pre-edge and scaling so that the post-edge L-edge absorption length was equal to that provided by  the Center for X-ray Optics (CXRO), 1.81$\times$10$^4$ (g cm$^{-2}$)$^{-1}$ at at $\sim$730 eV.

\begin{figure}
 \vspace*{-0in}
\includegraphics[width=10.0in]
{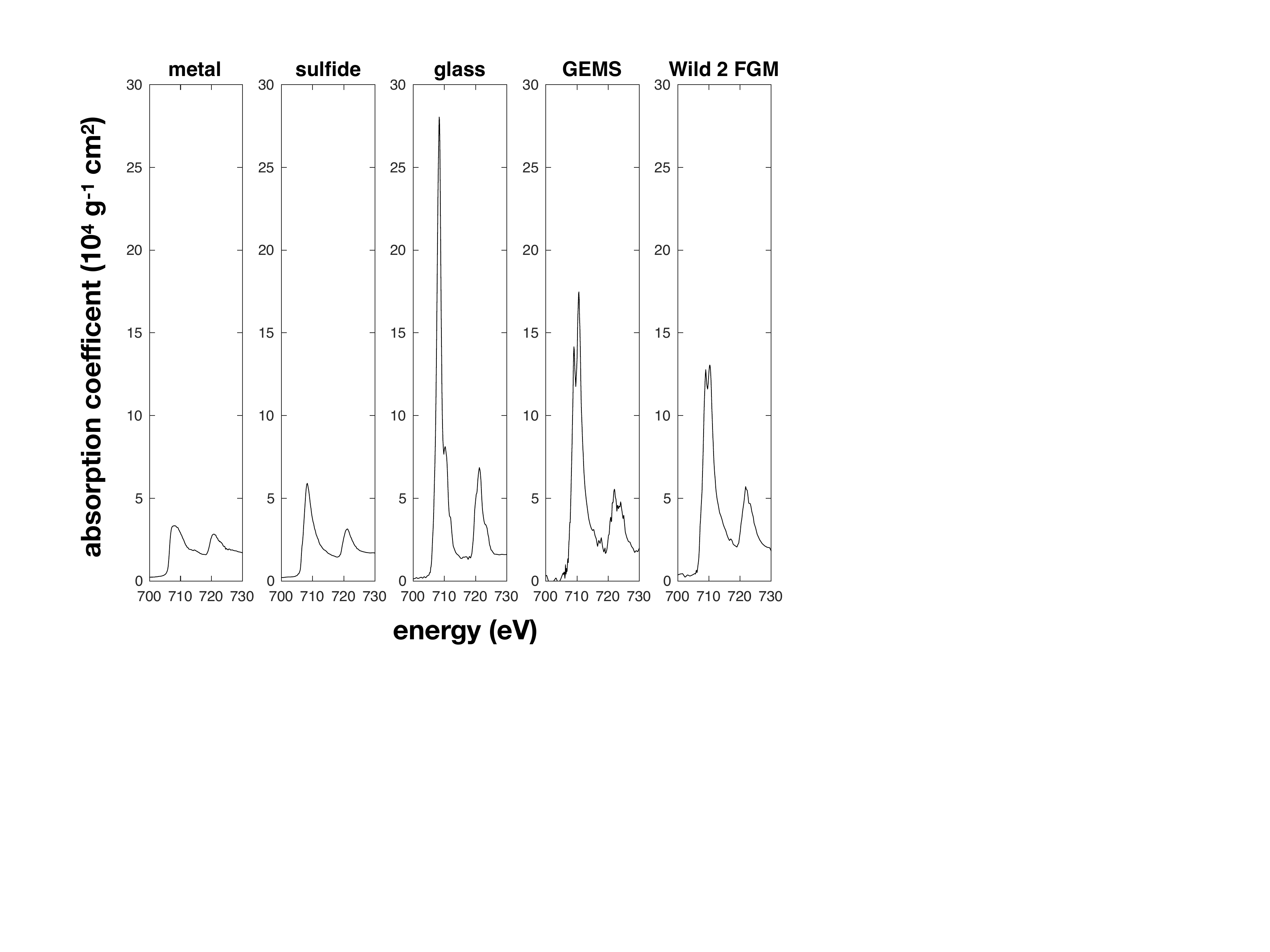}
 \vspace*{-2.5in}
\caption{Relevant spectra from our library of  Fe L-edge spectra of mineral standards.  Each absorption spectrum is normalized to 1.8$\times$10$^{4} $(g/cm$^{-2}$)$^{-1}$, the amplitude of the absorption step at $\sim$730 eV. }
\label{fig:spectra}
\end{figure}

Spectra that we used for this study are shown in Fig. \ref{fig:spectra}. 
For fitting astronomical data,  we used three standards:  metal, troilite (FeS) and, because infrared observations of the ISM  indicate that $>$97\% of ISM silicates are amorphous \citep{kemper2005}, we used Fe-bearing volcanic reticulite glass, an amorphous silicate.   
 Although the spectral shapes are similar, the absorption strengths of the resonant peaks relative to the tail of the normalized spectra are highly variable.    Metal shows the smallest amplitude within resonant peaks while silicate glass shows the largest amplitude.
 
 The absorption cross-section as a function of energy, $\sigma(E)$, depends on the overlap of the pre-excitation and post-excitation wavefunctions:

\begin{equation}
	\sigma(E) \propto \left< \psi_f | T | \psi_i \right>
\end{equation}
 
where $T$ is commonly the dipole operator, $\ket{\psi_i}$ is the wavefunction of the core electron before absorbing the X-ray, and $\ket{\psi_f}$ is the wavefunction of the excited state.  
If the overlap of the two wavefunctions is small then $\sigma(E)$ will be small, all other factors held constant.  
In the case of a conductive material such as Fe metal or sulfide, the excited state is often located in a conductive band and therefore delocalized which reduces the overlap $\bra{\psi_f} T \ket{\psi_i}$ and hence suppresses the intensity of the resonance peak.  
An ionic solid such as FeO has a more localized excited state and therefore a larger overlap and more intense resonance peak.  
In general, there are many factors determining $\ket{\psi_f}$, many-electron interactions are often important, and in some cases a simple dipole transition is not sufficient to describe the absorption.  
Room temperature magnetite provides an example of such an exception \citep{Friak}. 
The band structure is insulating for the majority-spin electrons ($e_g$ symmetry), yet conductive for minority-spin electrons ($t_{2g}$ symmetry) and it has an intensity intermediate between Fe metal and silicates.  
Generally the peak intensity is a complex emergent property depending on the details of the wavefunctions and transition and would have to be measured or simulated for each specific material.
Therefore, we have restricted our investigation to spectra which represent candidate ISM phases and within this set of materials, oxidized iron invariably has a greater peak amplitude than reduced iron.

\begin{figure}
 \vspace*{-0.4in}
\includegraphics[width=10.0in]{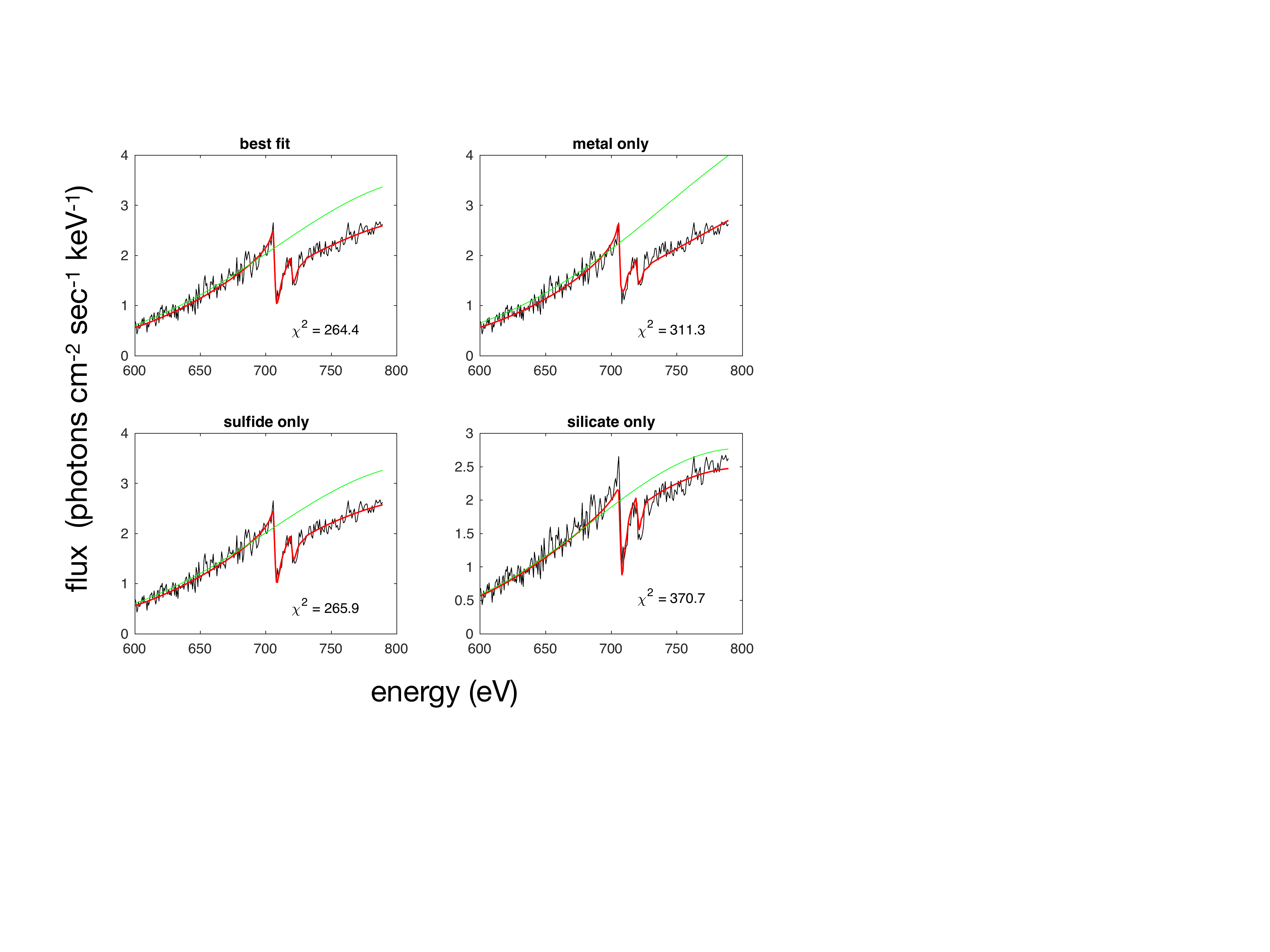}
 \vspace*{-2.0in}
\caption{Best fit to Cyg X-1 absorption data using all three components, and the best fits using only single components (metal, sulfide, and silicate).  The green curves are the continuum components of each fit. There are 241 degrees of freedom in the best fit, and 243 degrees of freedom in the other three.  Data were rebinned from the original dataset by a factor of four.
}
\label{fig:fits}
\end{figure}

\begin{figure}
\includegraphics[width=7in]{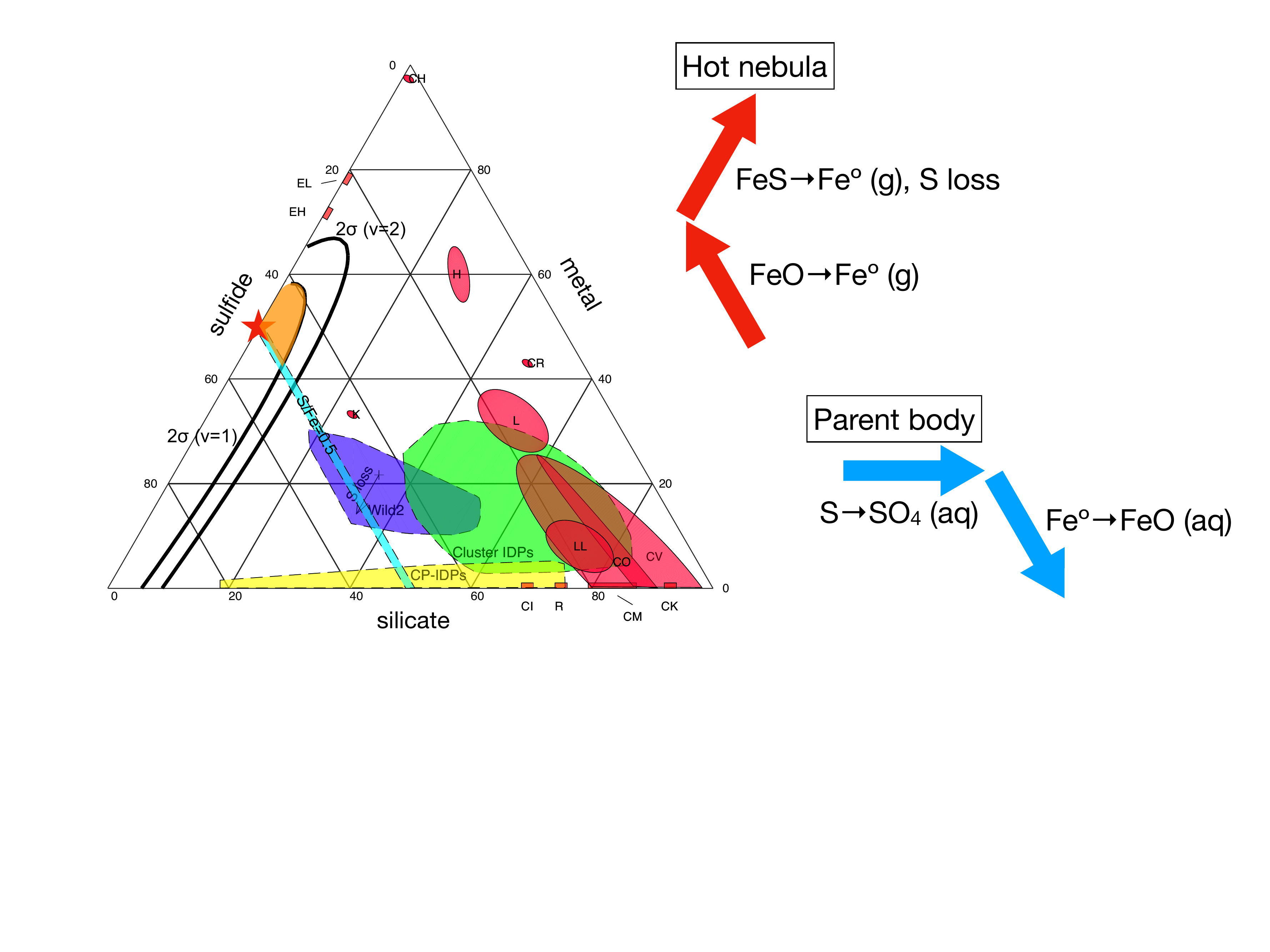}
 \vspace*{-1.8in}
\caption{(Left) Oxidation state of Fe in extraterrestrial materials. ISM measurement, reported here, is shown as $\Delta\chi^2=4$ (2$\sigma$, $\nu=1$)  and $\Delta\chi^2=6.13$ (2$\sigma$, $\nu=2$) contours over the space of metal,  sulfide and silicate fractions.  The contour is projected onto the compositional ternary from the third dimension, $a_{\rm cutoff}$. The (2$\sigma$, $\nu=1$) contour projects to $2\sigma$ confidence intervals on each axis.   The expectation value of a high-temperature condensation sequence with  solar abundances is shown by the red star.  Fe oxidation state  in major meteorite families (H, L, LL, EH, ...) and cometary materials (Wild 2, CP-IDPs \citep{westphal2009, ogliore2010}, cluster IDPs \citep{westphal2017}) are also shown.  If S/Fe=0.5 in the ISM, and S is bound to Fe in FeS (troilite), then this puts an additional constraint on the Fe oxidation state, leaving the orange shaded region bounded by the S/Fe=0.5 line and the 2$\sigma$ contour from {\em Chandra} observations (orange).  (Right) Direction of candidate oxidation and reduction pathways in the early Solar System.}
\label{fig:ox}
\end{figure}

\section{Analysis} \label{sec:analysis}

\subsection{Derivation of extinction spectra} \label{sec:absspectra}

Our astronomical X-ray observations consist of extinction   spectra, which are the result of both absorption and scattering. As pointed out by \citep{corrales2016}, astronomical X-ray Mie scattering cross-sections, and therefore extinction spectra, are sensitive to the particle size distribution.  The size distribution of the Fe-bearing phases within interstellar dust particles is not known. To account for this uncertainty we use the \citet{weingartner2001} dust size distribution, but introduce 
an upper size cutoff ($a_{\rm cutoff}$ in nm) as an additional free parameter.  (The \citet{weingartner2001} models are based on microwave and infrared emission, and optical and ultraviolet extinction, which are relatively insensitive to the internal structure of submicron dust particles.  However, Mie scattering of X-rays at the Fe L-edge ($\lambda = 1.7$\,nm) depends sensitively on the size of the scattering particles and their spatial distribution.  For example,  
Fe-bearing particles assembled in randomly-distributed, fractal-like aggregates would be indistinguishable in  X-ray observations from the same particles dispersed in the ISM as physically separated objects.)  
  To calculate scattering cross sections, we followed  \citet{draine2003} and \citet{corrales2016}.  We first calculated optical constants from synchrotron absorption data using the Kramers-Kronig relation implemented in kkcalc (v0.7.3), a Python package written and made publicly available by Ben Watts and Dan Lauk.    We then calculated the total scattering spectrum by integrating over the Draine WD01 \citep{weingartner2001} interstellar dust size spectrum, with an upper size cutoff left as a free parameter.   As a check, we reproduced the published extinction spectra in \citet{corrales2016} in their published test case.

\subsection{Spectral fitting to expected phases} \label{sec:absspectra}

We fit the astronomical data against derived extinction spectra from  metal, sulfide and amorphous silicate using a 7 parameter least squares fit.  We used a quadratic continuum with three parameters, three more for the metal/sulfide/oxide amplitudes, and $a_{\rm cutoff}$.  The best fit can be seen in Fig. \ref{fig:fits}. 
$\chi^2$ was 264.4 with 241 degrees of freedom at the best fit.  Here we used statistical errors only.  The small excess in $\chi^2$ may be due to uncertainties in the instrumental response matrix.  The optimum upper size cutoff $a_{\rm cutoff}$ was significantly below the WD01 upper size limit $165\pm25$\,nm (1$\sigma$).  

In Fig. \ref{fig:ox} we show the confidence region for oxidation state based on running multiple fits. The {\em Chandra} data are inconsistent with a large fraction of Fe in the silicate phase.  We place a 2$\sigma$ upper limit on the fraction of Fe in silicate of 0.105. 

The  S/Fe ratio in the Solar System is $\sim$0.5.   If Fe and S are entirely condensed into Fe metal and FeS, then Fe should be  equally partitioned into metal and sulfide.  This value would be expected in a high-temperature condensation sequence,  in which Fe metal condenses first at  $\sim$1350K, followed by sulfidation of approximately half of the Fe metal at $\sim$700K.  
 This point is marked with the red star in Fig. \ref{fig:ox}.  Sulfidation of Fe would place an additional constraint on the oxidation state of Fe: the region of oxidation state to the left of the S/Fe = 0.5 is excluded. To the extent that sulfide is pyrrhotite (Fe$_{1-x}$S, $x\le0.2$), or that sulfur condenses in significant quantities in other phases --- H$_2$S or polycyclic aromatic hydrocarbons (PAHs) --- this boundary will shift away from sulfide-rich composition in Fig. \ref{fig:ox}.

\subsection{GEMS and cometary fine-grained material} \label{sec:gems}

GEMS (Glass with Embedded Metals and Sulfides) constitute a major but enigmatic component of chondritic-porous interplanetary dust particles \citep{bradley}. They show anhedral morphology, and consist of nanophase metal inclusions in an amorphous silicate matrix, with iron sulfides either decorating the surfaces or, rarely, inside the particle. They show a restricted size range, between 100\,nm and 500\,nm in largest dimension. GEMS have been variously proposed to have a solar-system \citep{kellermessenger} or interstellar \citep{ishiiGEMS} origin.   Along with equilibrated aggregates, GEMS dominate the silicate fraction of CP-IDPs.   Here we test the hypothesis that interstellar silicates consist of GEMS as seen in CP-IDPs. 

We used the average of two absorption spectra of GEMS acquired at the ALS from CP-IDP RB-12A31-2 (Fig. \ref{fig:STXM}B,C).  We then computed the extinction spectrum using the method described in  \ref{sec:absspectra}, but rather than integrating over a truncated \citet{weingartner2001} size spectrum, we summed over the  sizes reported by  \citet{kellermessenger} in analyses of 44 GEMS.  Because the Keller \& Messenger size distribution is dominated by large particles, the scattering cross-section is larger.  We then generated a synthetic absorption spectrum, using the statistical noise  of the Cyg X-1 spectrum (Fig. \ref{fig:fits}), but with the GEMS extinction spectrum applied to a straight-line continuum, and fit this to the metal, sulfide and amorphous silicate components as before.  The fit is poor, with $\chi^2 = 372$ for 241 degrees of freedom.  We note that this CP-IDP, like many, shows some signatures of oxidation during atmospheric entry, so it is possible that some Fe, originally in metal and sulfide, oxidized and went into the silicate.  This process, if it occurred, does not appear to have   removed significant amounts of metals and sulfides from the particles (Fig. \ref{fig:STXM}).   We address this issue in the discussion.

We also compared the {\em Chandra} data extinction spectra derived from analyses of fine-grained material from a particle (``Andromeda'', \citealt{gainsforth2016}) collected and returned by the Stardust spacecraft from the coma of comet Wild 2.  We followed the same procedure outlined above, but used the non-truncated WD01 size distribution.  The fit is similarly poor ($\chi^2$ = 412, 241 d.o.f.).  

\section{Discussion} \label{sec:discussion}

Our measurement of the Fe oxidation state of the ISM is inconsistent with a major component of Fe in silicate, and is broadly consistent with the lack of correlation of Fe with Mg and Si reported by \citet{dwek2016}.  The best fit to the data indicate that no more than 8\% (2$\sigma$) of Fe in the ISM, at least in this line of sight, resides in silicate.  Whether or not this is consistent with the gas-phase depletions reported by \citet{dwek2016} is an open question.

Iron in the interstellar medium is strongly reduced by comparison with most, but not all, materials from small bodies in the Solar System (Fig. \ref{fig:ox}).  The oxidation state observed in laboratory analyses was determined by the oxygen fugacity in the formation region of these materials, and by oxidation, often due to aqueous chemistry, after incorporation into parent bodies.  Carbonaceous chondrites (CI, CM, CK, CV, CO, CR) are generally the most oxidized.   While L ordinary chondrites (OCs) lie near the carbonaceous chondrites in oxidation state, H OCs are more metal-rich.  No meteoritic or cometary oxidation state overlaps with the 2$\sigma$ confidence region of the ISM. 

\subsection{Redox reactions in the early Solar System}

This paper does not address the complexity of processes that converted presumably well-mixed, homogeneous ISM material into the diversity of small bodies in the early Solar System.  We observe only the initial and final states.  However, it is clear that net oxidation, reduction and loss of sulfur (at least in the form of FeS) are required in a variety of combinations to arrive at the remarkable diversity of oxidation states shown in Fig. \ref{fig:ox}. 

A hot gas with cosmic abundances is highly reducing because of the extreme excess of H.   ISM dust evaporated and recondensed in such a gas would be expected to have an oxidation state indicated by the star shown in Fig \ref{fig:ox}.   In a cooling gas,  S does not go into solids until quite late, when S sulfidizes Fe metal at $\sim$700K, so if gas is lost from the system, for example by stellar outflows or out-of-plane outward transport from the inner Solar System, it  preferentially removes S.  As a consequence, less FeS is eventually produced.  Similarly, FeS might be lost through relatively mild heating and conversion of FeS to Fe metal, with loss of S from the large-scale transport.

A potential net pathway from the ISM to the enstatite  (EH and EL, Fig. 4) chondrites, then, may have been S loss in a reducing hot nebula of near cosmic abundance, with abundant H.   This is perhaps consistent with the probable formation of EH and EL materials in the inner solar system. 
Another possible mechanism is energetic ion irradiation, although this is probably ruled out as a major process responsible for the oxidation state of enstatite chondrites because it would have led to strong and obvious isotopic anomalies.

The net pathway for carbonaceous chondrites and ordinary chondrites, except perhaps for H chondrites, may have been oxidation of metal, as well as oxidation of sulfur to deplete sulfides in favor of sulfates.   The most likely candidate for these processes is in an aqueous environment in a wet parent body.  This may be consistent with the proposed origin of the carbonaceous chondrites at or just beyond the snowline in the early Solar System.

The Fe oxidation state in the ISM and that of cometary material is disjoint.  The path from the ISM to cometary materials remains unclear, since comets were probably too cold to have ever hosted liquid water.   

An unsolved problem in meteoritics is an understanding of the origin of Fe-rich olivine (Fe$_x$Mg$_{2-x}$SiO$_4$) and low-Ca pyroxene  Fe$_x$Mg$_{1-x}$SiO$_3$, which are unexpected to condense from a gas cooling from high temperature \citep{grossman}.   The Fe content of these minerals is an increasing function of oxidation state.   Even in an oxygen-rich condensing gas, these minerals are expected to be very Fe-poor, with fayalitic contents (=Fe/(Fe+Mg)) of no more than 4\% \cite{grossman}. Before the return of samples from comet Wild 2, it was thought that oxidation of crystalline silicates might have occurred within the parent bodies of meteorites, during aqueous alteration or thermal metamorphism.  However, olivines returned from comet Wild 2 show a very broad and nearly flat distribution in fayalitic content from 0 to $\sim$40\%, and a range from 0\% to 97\% \citep{frank}, so the oxidation appears to have taken place in a nebular environment, through an unknown mechanism. 

We note that all three collections of cometary materials have the potential to have suffered oxidation state change, or the appearance of oxidation state change through sample bias, between release from the parent body and analysis in the laboratory. Samples from comet Wild 2 were captured in the coma at 6.1 km/sec, which may have led to reduction of Fe through smelting:  the arrow in Fig. \ref{fig:ox} indicates that the original material may have been less reduced than observed in laboratory analyses.  Cometary materials captured in the stratosphere may suffer from biases against metal-rich, high-density particles, because of their shorter residence time.  This is consistent with the metal-poor composition of individual small CP-IDPS, in contrast to the relatively metal-rich composition of particles from giant cluster IDPs.
However, none of these capture or sample bias effects go in a direction to reconcile cometary oxidation state with that of the ISM.

\subsection{Are interstellar silicates unoxidized GEMS-like materials?}

The GEMS Fe L-edge absorption spectrum (Fig. \ref{fig:spectra}) shows an enhanced resonance consistent with a major fraction of Fe in the silicate phase. It is possible that this oxidation of Fe could occur at some time between residence in the ISM and analysis in the laboratory --  in the solar nebula, during atmospheric entry, or due to exposure to terrestial oxygen.   To test the hypothesis that interstellar Fe resides in GEMS-like silicates but in which all the Fe is sequestered in metals and sulfides, we measured the size distribution of metals and sulfides in a GEMS, then calculated a synthetic extinction spectrum using the measured metal and sulfide absorption spectrum and the observed size distribution of metals and sulfides.  We assumed equal amounts of metal and sulfide, as expected by condensation from cosmic abundances. Unsurprisingly, because of the small size of the particles, the synthetic extinction spectrum did not exhibit a significant pre-edge scattering feature --- the turn-up in the data between 680 eV and 705 eV in Fig. \ref{fig:fits} ---  so the fit to the {\em Chandra} data was poor, with $\chi^2 = 313$, indicating that  GEMS in which Fe is entirely sequestered in metal and sulfides are an unlikely candidate for interstellar silicates.

\subsection{Summary}

A comparison of Fe L-edge absorption spectrum of continuum X-rays from Cygnus X-1 with extinction spectra derived from synchrotron X-ray microprobe analyses of mineral standards leads to the following conclusions:

$\bullet$  Observations are consistent with the hypothesis that Fe is sequestered principally in sulfides and metals,
at least in this line of sight. The data are consistent with Fe evenly divided between sulfide and metal, a composition that would be expected if  Fe and S reside entirely in metal and sulfide, e.g. in a high-temperature condensation sequence.  Less than 8\% (2$\sigma$) of Fe resides in amorphous silicates.

$\bullet$ The observations are inconsistent with the hypothesis that interstellar silicates consist of GEMS as found in chondritic-porous IDPs.  This conclusion is robust even if some Fe is oxidized between residence in the ISM and observation in the laboratory. 

The same methodology may be applied to explore the oxidation state of ISM Fe in other lines of sight, toward other X-ray binaries.  

\section{Acknowledgments}

We thank Gary Huss for providing the reticulite glass standard and Young-Sang Yu for help acquiring  the spectrum at STXM ALS Beamline 11.0.2.2.
We thank Dave Joswiak for supplying IDP RB-12A31-2.
This work was supported by NASA grants NNX16AI15G and NNX17AF86G.
Work at the Molecular Foundry and Advanced Light Source was supported by the Office of Science, Office of Basic Energy Sciences, of the U.S. Department of Energy under Contract No. DE-AC02-05CH11231.



\end{document}